# The Literature Review Network: An Explainable Artificial Intelligence for Systematic Literature Reviews, Meta-analyses, and Method Development.

[1]Joshua Morriss PhD*, [2,3]Tod Brindle PhD MSN RN CWCN*, [4]Jessica Bah Rösman PhD, [1]Daniel Reibsamen BS, [5,6]Andreas Enz MD

*co-first authors

## Abstract

Systematic literature reviews are the highest quality of evidence in research. However, the review process is hindered by significant resource and data constraints. The Literature Review Network (LRN) is the first of its kind explainable AI platform adhering to PRISMA 2020 standards, designed to automate the entire literature review process. LRN was evaluated in the domain of surgical glove practices using 3 search strings developed by experts to query PubMed. A non-expert trained all LRN models. Performance was benchmarked against an expert manual review. Explainability and performance metrics assessed LRN's ability to replicate the experts' review. Concordance was measured with the Jaccard index and confusion matrices. Researchers were blinded to the other's results until study completion. Overlapping studies were integrated into an LRN-generated systematic review. LRN models demonstrated superior classification accuracy without expert training, achieving 84.78% and 85.71% accuracy. The highest performance model achieved high interrater reliability ($\kappa = 0.4953$) and explainability metrics, linking 'reduce', 'accident', and 'sharp' with 'double-gloving'. Another LRN model covered 91.51% of the relevant literature despite diverging from the non-expert's judgments ($\kappa = 0.2174$), with the terms 'latex', 'double' (gloves), and 'indication'. LRN outperformed the manual review (19,920 minutes over 11 months), reducing the entire process to 288.6 minutes over 5 days. This study demonstrates that explainable AI does not require expert training to successfully conduct PRISMA-compliant systematic literature reviews like an expert. LRN summarized the results of surgical glove studies and identified themes that were nearly identical to the clinical researchers' findings. Explainable AI can accurately expedite our understanding of clinical practices, potentially revolutionizing healthcare research.

## I. Author Affiliations

**1** Ziplitics, Inc., Midlothian, USA
**2** Mölnlycke Health Care, Norcross, USA
**3** Convatec LTD, Bridgewater, USA
**4** Mölnlycke Health Care, Göteborg, SE
**5** Helios Kliniken Schwerin, Schwerin, DEU
**6** Universitätsmedizin Rostock, Rostock, DEU

## 1. Introduction

During surgery, the aseptic barrier exists as the primary method to protect the operating room personnel from the patient and the patient from the clinical team. As part of what is known today as personal protective equipment (PPE), gloves provide the aseptic barrier and protection of the hands from potentially infectious materials being cross contaminated within the surgical field. However, despite the innovations seen in surgical glove raw materials and manufacturing, surgeons and nurses have known that glove damage during surgery occurs[1], [2].In fact, for over seven decades clinicians have reported and studied glove damage in surgery and have described from large, visible tears to microperforations, not visible to the human eye[3], [4], [5].

Nothing is more feared in surgical practice than the surgical site infection. The increase in morbidity and mortality for the patient, in addition to the increases in costs for the health care system, are well described in the literature. While glove damage during surgery is nearly impossible to link as a direct cause of surgical site infection due to a extensive list of confounders, the risk of infection causes surgical teams to take every precaution to decrease infection. Additionally, provider safety from contracting infectious disease was highlighted during historical pandemics such as HIV-Aids in the 1980s[6], [7] and recently with SARS-COV-19[8], [9], [10]. Additionally, more common viral pathogen such as hepatitis B and C are inherent risks for the reported 400,000 sharp injuries estimated per year[11]. Therefore, research such literature reviews are needed to understand the incidence, prevalence, time-lapse and potential causes of glove damage during surgery to improve the safety of the provider and patient.

Systematic literature reviews (SLRs) are the gold standard in clinical and preclinical research, informing public policy, clinical guidelines, and R&D for medical devices and pharmaceuticals[12], [13]. In clinical practice, the use of systematic reviews and metanalysis guide the development of clinical practice guidelines, which direct practice change to achieve the best outcomes. Despite their importance, producing SLRs is challenging due to the sheer volume of research published annually, estimated at >1 million studies, leading to reporting biases and gaps in evidence[14], [15]. Furthermore, SLRs incur substantial financial burdens, with an average expenditure of approximately $141,194.80 per SLR, and a significant time investment of approximately 1.72 years per researcher[16]. Advances in natural language processing (NLP) and machine learning (ML), alongside web-based large language model (LLM) and artificial intelligence (AI) platforms, have been proposed to streamline the SLR process. These technologies aim to enhance data extraction and text classification, but they often fall short in automation, continuous updateability, and particularly in explainability—the ability for humans to understand and trust the decisions made by an AI to achieve its outputs[17], [18]. Current NLP-ML and AI solutions lack automation and explainability, and do not meet the rigorous standards of high-quality research frameworks like Preferred Reporting Items for Systematic Reviews and Meta-Analyses (PRISMA) 2020[19]. This apparent lack of transparent and explainable processes in NLP-ML and AI applications for SLRs has notably contributed to hesitancy in their widespread adoption for clinical and preclinical research.

Addressing these concerns is the Literature Review Network (LRN), an innovative explainable AI (XAI) platform designed for SLRs, meta-analyses, and real-world data research. This study evaluated LRN's effectiveness by comparing its accuracy and reliability to a traditional, human-conducted SLR. Specifically, we assessed LRN's ability to achieve high overall model accuracy and interrater reliability, measured by Cohen's kappa, with fewer iterations. To evaluate this XAI's accuracy, our reference was a human SLR conducted by three subject matter expert (SME) researchers and six reviewers focusing on surgical glove damage and change frequency, completed without NLP-ML or AI assistance. This manual SLR was part of a larger study including 4 SLRs conducted in parallel, with the objective being to determine the best available evidence to describe four key fundamental principles of surgical gloving practice: glove fit, double gloving, puncture indication, and glove change frequency. Additionally, the previous study queried clinical literature with multiple search strings and documented different search strategies. Therefore, the second aim of this study was to determine if an XAI could streamline the SLR process by finding similar insights on surgical gloving practice with fewer searches compared to conventional methods. This comparison aims to determine whether LRN can streamline the SLR process, achieving comparable insights with fewer searches and in less time. Lastly, this study aimed to evaluate an AI's capability to summarize the literature with little to no instruction by an SME, approximating themes like those identified by SMEs. A qualitative assessment of the LRN-generated SLR was done by this study's surgical glove SMEs (TB, AE, JBR) to determine the similarities and differences between the human-identified and XAI-identified themes.

# 2. Methods

## 2.1 Search Strategy, Data Extraction, Review Procedures

Based on the protocol and PICOT framework used for the human SLR, three human SMEs (TB, AE, JBR) initially formulated three separate search query strings that were used in this study. These search strings were then converted into LRN queries, as recorded in **Table 1**, and a set of concept rules which served as the basis for LRN's reinforcement learning. Search strings were subjected to LRN version 2.0 (LRN v2.0), which employed a word embedding model that mapped concepts with the Unified Medical Language System (UMLS) Metathesaurus[20]. LRN processed each of these unique queries independently as three LRN models. These LRN models were configured to query the PubMed database for relevant literature, utilizing the PubMed API for data retrieval. Of the 262 studies that were identified and included in the manual SLR by the SMEs, the PubMed ID (PMID) was retrieved for only 212 studies. LRN's screening mechanism involved the automatic ineligibility of records that either lacked an abstract, were published in Russian or Chinese, or were identified as duplicates. Russian and Chinese studies were excluded due to the former being a low-resource language, while the later presented complications related to accurate word segmentation with LRN[21]. Records that met the exclusion criteria were also automatically excluded by LRN (**Table 1**), and composed an out-of-domain or negative dataset to train LRN's discriminative algorithms[22]. Levels of evidence considered were randomized controlled trials, cohort studies, observational studies (retrospective and prospective), quasi-experimental studies, SLRs, and meta-analyses.

LRN v2.0 operated within a reinforcement learning with human feedback (RLHF) framework. A non-SME in surgical gloving practice (JM) was responsible for training these three LRN models. When configuring this model and deriving the initial concept ruleset, the non-SME (JM) adhered to the protocol from the manual SLRs. The non-SME began the iterative learning process by first defining language rules associated with the INCLUDE and EXCLUDE classes (**Table 2**). Studies were then classified as either INCLUDE or EXCLUDE based on the non-SME's feedback and LRN's decisions. A RLHF loop initiated where LRN presented its findings with 20 labeled records. Per each iteration, the non-SME reviewed the performance metrics, word cloud, and correlation and coverage tables to assess LRN's associations. Based on these associations, the non-SME modified the concept ruleset to add or remove rules. The non-SME then screened each record's title and abstract, and provided feedback by assigning a label (e.g., INCLUDE or EXCLUDE). This assigned label was compared with LRN's predicted classification label. For this literature review, each LRN model was trained for 4 total iterations, or 3 RLHF iterations.

## 2.2 Explainable Artificial Intelligence Framework for Research

LRN utilized a combination of a metaheuristic wrapper, weak supervision models, and discriminative algorithms. LRN first extracted natural language features from the corpus, with the wrapper optimizing feature selection for semantic analysis considering user-defined language rules. This metaheuristic wrapper feature selection technique reduced the complex feature space inherent to natural language data[23]. Weak supervision models in LRN operated under a matrix completion methodology and generated several rudimentary models from the concept ruleset. These models, despite their inaccuracies, can effectively label unstructured literature[24].

Discriminative models refined labels by analyzing consensus and discrepancies among weak models, effectively handling correlated labels from weak supervision sources without requiring labeled data[24]. Performance metrics of recall, precision, and F-score were automatically calculated by LRN for each label, in addition to the overall model accuracy and Cohen's kappa. Additionally, LRN calculated potential scores for each record to balance exploration of new linguistic models and exploitation of established models and data structures for classification[25]. During RLHF, LRN presented 20 records with the highest potential score for feedback. Data visualizations and correlation tables were produced to clarify literature screening decisions for each iteration. The relationships between generative AI parameters were quantitatively assessed using Pearson's chi-squared test, adjusted by Cramer's V, and corrected for significance using the Benjamini-Hochberg method[26], [27], [28]. A LRN "AI Package Insert" for each search string documented all metrics and model decision-making processes. The highest performance model was identified by superior Cohen's kappa and accuracy, while the optimally balanced model was determined by the optimal balance of true positives to false negatives across search strings and iterations.

## 2.3 Analysis of LRN Search Alignment with SME Review

To critically appraise each search strategy and determine if LRN could effectively streamline the SLR process, the similarity was assessed between the three search strings and the SME-curated library. The unique reports from all three search strings were first pooled into a single corpus and deduplicated. The model with the highest Cohen's kappa and overall accuracy from each search string was then used to classify the entire corpus. Non-SME feedback from all the models, which were the non-SME assigned labels for unique records, was incorporated into the dataset for classification by each string's optimal model. PMIDs were used as the identifiers for comparing the SME-curated library and the LRN model classifications. Three sets of PMIDs corresponded with the three optimal models from each search string. Concordance between each LRN model's predictions and the SME selections was done through the calculation of a Jaccard index for quantifying the overlap between studies classified as "INCLUDE" by each LRN model and those identified in the manual SLR[29]. Bootstrapping for 1 million replications was performed to determine the significance level of the similarity between the sets of PMIDs and the SME library. A Jaccard index was also calculated comparing each search string against the others to determine if there was significant overlap between, for example, search string 1 and search string 2, search string 1 and search string 3, and search string 2 and search string 3. P-values were adjusted using the Benjamini-Hochberg method[26]. Confusion matrices further explored the performance of the optimal models for each string in classifying the literature, providing coverage statistics on true positive (INCLUDE) and true negative (EXCLUDE) studies, as well as false positive and false negative cases.

## 2.4 Text Summarization via LRN written SLRs

In alignment with PRISMA 2020 guidelines, the optimally balanced model across all search strings was applied across the three search strings to classify the surgical glove corpus. This model automatically generated a PRISMA 2020 flow diagram detailing study identification, screening, and inclusion processes[30]. Classified 'INCLUDE' studies were prepared for summarization using an LLM that was based on OpenAI's GPT-4-turbo[31]. LRN identified, labeled, and embedded full-text reports, which were then subjected to a series of user questions (**Table 3**). Full-text reports were formatted and embedded for further processing, utilizing Langchain's Retrieval-Augmented Generation (RAG) technique to select relevant token segments, likelihood of data leakage and hallucinations was reduced[32]. LRN's LLM merged its outputs into a single document, encompassing a LRN-generated SLR. Time to complete draft metric for the manual SLR quantified the cumulative human labor hours required for identifying, screening, assessing eligibility, and incorporating full-text reports. In contrast, this metric for the LRN models considered the total human labor hours needed to configure the three models (iteration 1), complete the RLHF iterations for each model (iterations 2+), and the computation times for each LRN model to complete an iteration. Lastly, the three SMEs (TB, AE, JBR) reviewed the LRN-generated SLR to determine thematic similarities and differences.

## 3. Results

### 3.1 LRN Model Performance Metrics

Optimal performance was achieved with search string 3, iteration 3, which yielded the highest performance model with an overall accuracy of 84.78% and a Cohen's kappa of 0.4953 (**Table 4**). This contrasts with the best-performing models from search string 1 iteration 2 and search string 2 iteration 3, which produced Cohen's kappa values of 0.2174 and 0.0183 and overall accuracies of 85.71% and 58.62%, respectively. While string 1 engendered a model with 0.93% higher overall model accuracy than string 3, the EXCLUDE class performance metrics for string 1 were suboptimal compared to string 3 (**Tables 5-6**). Improvements in precision and recall for the EXCLUDE class in the LRN model for string 3 iteration 3 demonstrated the model's efficacy in filtering studies unrelated to the research aims, yet this was with a slight decrease in precision for the INCLUDE class (**Table 5**). Despite string 3's superior performance, a notable decrease in average potential from 85.44% to 49.28% was observed, suggesting a reduction in semantic information retention across iterations for the string 3 model (**Table 4**). Four iterations were executed for all three models, with non-SME assigned labels for 92 unique records. However, by the fourth iteration, all models exhibited signs of underfitting with a set of new rules added by the user. Therefore, each model retained its learned associations up to the third iteration. Initially, search string 3 identified 284 potential studies for inclusion, with the validated optimal model ultimately selecting 149 full-text reports from that subset of the literature (**Figure 1**). Across the three search strings, a total of 810 studies were initially identified as candidates for inclusion. From this study, the highest performance model (search string 3 iteration 3) was discovered to not be the optimally balanced model for classifying the literature. Instead, the model from search string 1 iteration 2 was determined to be the optimally balanced model. Thus, upon application of the optimal model from search string 1 iteration 2 to these records, 757 full-text reports were classified as 'INCLUDE' (**Figure 2**).

As an XAI, LRN models elucidate their inclusion and exclusion decisions for the literature through visual representations (**Figure 3**) and detailed quantitative tables. These correlations guided LRN's decision-making process. The highest performance model identified novel concepts in studies classified for inclusion, featuring terms such as 'reduce', 'accident', 'sharp', and 'double-gloving.' Comparatively, the concepts uncovered by the optimally balanced model during RLHF are also depicted in **Figure 3**. Leveraging RLHF, LRN ingests human feedback through natural language rules and establishes semantic contexts for these rules by establishing correlations between two unique concepts or numerical measures. Specifically, when two unique rules exhibited a high correlation, it indicated that the LRN model had contextually associated these terms as either co-occurring or related within the literature. The highest performance model (search string 3 iteration 3) identified significant correlations between concepts such as ‘nitrile’ and ‘examination glove(s)’ ($r = 0.601$, p-value = 5.302E-13), ‘condom’ and ‘hand washing’ ($r = 0.505$, p-value = 3.716E-09), ‘polychloroprene’ and ‘nitrile’ ($r = 0.495$, p-value = 8.080E-09), and ‘operation’ and ‘surgical glove(s)’ ($r = 0.490$, p-value = 1.052E-08). Additional significant correlations (p-value > 0.05) are documented in **Table 7**. For search string 1, the optimally balanced model identified significantly correlated rules such as ‘talc’ and ‘animals’ ($r = 0.404$, p-value = 2.424E-12), ‘condom’ and ‘antibiotic prophylaxis’ ($r = 0.397$, p-value = 6.424E-12), ‘exam glove’ and ‘examination glove’ ($r = 0.369$, p-value = 2.710E-10), and ‘operation’ and ‘surgical gloves’ ($r = 0.369$, p-value = 2.710E-10). Based on the performance of the LRN model derived from string 1 iteration 2 (**Table 4, Table 6**), this model was considered as the optimally balanced model, and was later used to classify the surgical glove corpus. Statistically significant, pertinent rules and relationships guiding the optimally balanced model’s classifications are presented in **Table 8**. The optimal model from search string 2, which underperformed compared to the models from search strings 1 and 3, revealed correlations between ‘exam glove’ and ‘nitrile (glove)’ ($r$ = 0.495, p-value = 2.098E-11), ‘examination’ and ‘examination glove’ ($r = 0.482$, p-value = 8.306E-11), ‘maxillofacial’ and ‘mandibular’ ($r = 0.461$, p-value = 6.203E-10), and ‘polychloroprene’ and ‘exam glove’ ($r = 0.401$, p-value = 1.409E-07).

## 3.2 LRN Productivity Metrics

The optimal models for each search string were deployed on the total LRN corpus of 810 studies. Of these 810 studies, a total of 194 identified studies could be found overlapping with the manual SME-curated library of 262 studies, of which 212 studies had PMIDs. From the LRN corpus of 810 studies, search string 1 classified 757 full-text reports, search string 2 classified 389 full-text reports, and search string 3 classified 674 full-text reports as INCLUDE (**Figure 4**). Evaluation of the overlap between the three search strings, the manual SLR library revealed varying degrees of alignment: search string 2 exhibited the highest similarity (Jaccard index = 0.3151, p-value = 3.000E-5). Search string 1 (Jaccard index = 0.2503, p-value = 3.538E-03) and search string 3 (Jaccard index = 0.2238, p-value = 3.538E-03) demonstrated reduced similarity (**Table 9**). Moreover, significant similarity was observed among the LRN search strings themselves. The overlap between search string 1 and search string 3 was the highest (Jaccard index = 0.8609, p-value < 1.000E-20), achieving nearly similar coverage of the literature. Lastly, moderate similarity was seen between search string 2 and search string 3 (Jaccard index = 0.4682, p-value < 1.000E-20) (**Table 9**). Search string 2 resulted in the highest number of false negatives, followed by search string 3 and search string 1 (**Figure 4**). Of note, the optimally balanced model from search string 1 classified the highest number of true positives (n = 194) yet had the highest number of false positives (n = 563) compared to the highest performance model from search string 3 (**Figure 4**). As mentioned previously, search string 1 iteration 2 was used to classify the final set of studies, synthesized in the LRN-written SLR.

Total human labor time to conduct the complete manual SLR on surgical glove practices was 19,920 minutes, or 332 hours. The manual search and identification of the final list of abstracts occurred over 5 months, while the division and reading of a study's full texts, evidence table generation, data analysis, and the manuscript creation took approximately 6 months. Comparatively, total human labor time to complete all three LRN models, to perform all similar analyses, and to generate a manuscript was 288.6 minutes, or nearly 4.81 hours, over the span of 5 consecutive days. Computation time, separate from the human labor time, was 1810.7 minutes, or 30.18 hours, as itemized by search string by iteration in **Table 10**.

# 4. Discussion

## 4.1 Explainable AI Performed SLRs equivalent to Experts in Surgical Gloving Practice

An observably high inter-rater reliability score (**Table 4**) from the LRN model trained on search string 3 iteration 3 coupled to a high overall accuracy demonstrated that LRN had the capacity to behave like a SME in screening and classifying the literature on surgical glove procedures (**Figure 1**). This LRN model was identified as the highest performance model and achieved high INCLUDE precision and recall (**Table 5**) despite receiving the fewest number of rules compared against the other two search strings (**Table 2**). However, a strategic trade-off occurred for string 3 which improved the EXCLUDE class precision at the expense of reducing the precision for the INCLUDE class, a well-characterized relationship in document retrieval[33]. Still, the reduction in precision for the INCLUDE class was minor relative to the EXCLUDE label, in exchange for all metrics for the EXCLUDE class to increase from 0% (**Table 5**). This rebalance validated the robustness for LRN models to screen the literature.

Although search string 1 iteration 2 exhibited a marginally higher overall accuracy by 0.93% compared to that of search string 3 iteration 3, it was at the expense of a reduction in both precision for the INCLUDE class and recall for the EXCLUDE class (**Table 6**). If search string 1 were misattributed as the highest performance model due to its overall model accuracy alone, its low EXCLUDE recall would suggest that the deployed model would have failed to correctly exclude more studies based on the exclusion criteria[34]. Interestingly, this was observed with the higher count of false positives compared to the LRN model produced via search string 3 (**Figure 4**). This comparison of the tabulated performance metrics across different search strings highlighted the importance of not solely relying on accuracy. Alternative metrics like Cohen's kappa and potential provided deeper insights into a LRN model's alignment with user evaluations and its proficiency to extract relevant information from the literature.

LRN was initialized as a clean state XAI with the goal of maximizing overall accuracy, Cohen's kappa, and average potential for any given research objective. User feedback during configuration, in the form of natural language rules and inclusion or exclusion criteria from the non-SME (JM), was part of the RLHF framework that enabled an initialized LRN model to exploit information relevant to the research question. RLHF provided advantages such as the conservation of online computational resources by pre-computing optimal solutions offline based on incorporated feedback[35]. Conventional supervised ML models or supervised AI cannot exceed the performance of SMEs and can only mimic the behavior of an SME, as these systems learn from a training data set that is labeled by the SME[35]. This is in direct contrast to RLHF, which enabled LRN to exceed such benchmarks. Superior classification capabilities were observed for both search string 1 iteration 2 and search string 3 iteration 3, identifying key literature on surgical glove procedures (**Tables 4-6**). Of note was search string 3, which achieved high accuracy and high inter-rater reliability. The lack of SME training in all three models might have affected the interpretability of Cohen's kappa. Integrating additional well-calibrated rules for INCLUDE and EXCLUDE classes from the SME may have minimized the risk of underfitting for all models, a tendency observed by the fourth iteration; however, LRN's robustness was evident, with considerable accuracy achieved across search string 2 and search string 3. Remarkably, despite minimal alignment between non-SME user inputs and the LRN model classifications for string 2, a total of 194 out of 212 full-text reports were accurately identified as included studies within the SLR.

Comparing the outputs of the LRN model SLR with the results of the manual SLR on double-gloving resulted in the identification of a striking capability of this new tool for researchers. Even though the human interface for the LRN network (JM) was not a trained clinician, nor did the non-SME understood the depth of the clinical issue surrounding glove damage during surgery, the simplicity of marking LRN derived abstracts as INCLUDE/EXCLUDE based on the overarching research question, matched the efforts and knowledge of the SME. While the researchers of the original manual process had the advantage of clinical experience and knowledge of specific connections between materials and surgical behavior to guide their search strategies, this was over come through LRN's capability to identify relationships between terms in the literature and drawing inferences based on simple human feedback. The broader implications of these findings substantiate that SME training for LRN is not essential for LRN to achieve literature reviews equal in depth and accuracy of an SME.

## 4.2 Explainable AI Identified Key Themes equivalent to Experts in Surgical Gloving Practice

Pairing of non-SME provided rules included those such as 'nitrile' and 'examination glove(s)', both of which were classified by the non-SME and by LRN as EXCLUDE rules. The correlation between 'condom' and 'hand washing,' two EXCLUDE rules, presents an intriguing case; while individually relevant to latex material and practices associated with hygiene, respectively, their association was not directly pertinent to the core objectives of this study. These findings signify LRN's ability to discern and explicitly identify concepts deemed irrelevant to the study's objective. Moreover, the correlation between 'polychloroprene' and 'nitrile' across

INCLUDE and EXCLUDE categories, respectively, demonstrates LRN's nuanced comprehension of these materials as alternative options for surgical gloves and showcases LRN's mechanism for contextual differentiation. Furthermore, as presented through the visualization tools, the emphasis on 'reduce,' 'sharp,' 'contaminate,' 'tear,' 'reinforcement,' and 'double-gloving' as connected concepts and procedural measures for enhancing barrier protection between the clinician and the patient encapsulates LRN's successful extraction of critical procedural insights from the literature (**Figure 3**). These inclusion concepts' prominence within the model's classification decisions aligned closely with the study's investigative focus on surgical gloving practices. Through iterative feedback and rule adjustment, LRN demonstrated progressively enhanced specificity and reduced the inclusion of irrelevant studies.

Semantic understanding in both explicit and implicit contexts is necessary for an AI model to adequately capture the predicate-argument structures of human language. Significant development with respect to this area has been undertaken for different neural network architectures[36]. Moreover, AI language models that leverage context-sensitive features present valuable applications in the fields of preclinical and clinical research and development[37], [38], [39], [40]. Two modalities LRN leverages to explain its decision-making processes are tabular methods and advanced visualization tools. LRN is the first XAI capable of semantic understanding while writing SLRs, conducting meta-analyses, and method development in both explicit and implicit contexts as highlighted by the predicate-argument relationships between concepts in **Table 7** and **Table 8**.

After completion of the search methodology which accomplished the identification of the totality of articles selected through the original manual process, the final manuscripts were loaded into an LLM as a boundary condition and were asked to summarize the information (**Table 3**). Interestingly, three of the conclusions of the AI induced summary matched similar consensus statements derived from a group of surgeons and nurses completing the original systematic reviews. While not an original aim of this study, the XAI summary recommended the use of double gloving to reduce the risk of aseptic barrier breach, frequent changing of gloves during surgical procedures, especially in orthopedic surgery and that specific surgical procedures may require special considerations, such as suturing. These recommendations were like those proposed by experts in the field who had both practical personal experience as well as the benefit of reading the entire full text articles. Future studies using methods like BLEU should assess the accuracy of this XAI SLR compared to traditional methods[41].

## 4.3 Explainable AI Improved Productivity and Streamlined SLRs with Fewer Searches

In evaluating LRN's effectiveness to replicate the manually conducted SLRs, it was observed that 194 records from LRN search strings overlapped with the 262 reports included in the manual SLR. Interestingly, following full text review by the manual reviewers described in the ground truth SLR, the 262 reports were reduced to a final count of 165 by their specific inclusion and exclusion criteria. Thus, nearly all articles identified through a manual methodology including four databases (Pubmed, Embase, Google Scholar, Cochrane) were captured by LRN using PubMed alone. Search string 2 demonstrated the highest Jaccard index, which was attributed to a reduced number of false positives, yet it also exhibited the highest number of false negatives, missing 52 full-text reports (**Figure 2**). This discrepancy highlighted that search strings 1 and 3, showing a high degree of statistical similarity, could sufficiently cover the literature corpus for this study. Furthermore, search string 1 effectively covered 91.51% of the potential studies which overlapped with both LRN's corpus and the manual SLR. Taken together, LRN evidenced that the use of both search string 1 and search string 2 was unnecessary for this study (**Table 1**). In addition, due to the high similarity between search strings 1 and 3, it can be hypothesized that further refinement of either string's models would have rendered the use of one search string adequate. These results evince that the efficiency of LRN in literature identification, screening, and evaluation or inclusion (**Figure 4**) hinges on model training and validation, as evidenced by performance metrics, rather than the number of search strings employed (**Tables 4, 7-9**).

By design, LRN prioritized minimizing false negatives to ensure comprehensive capture of relevant documents. This approach demonstrated the potential of SME involvement in training LRN models to significantly reduce false positives. A major limitation of this study was overreliance on PubMed, which led to a reduction in the number of potential studies retrieved from 262 to 212, marking a 19.08% decrease. Furthermore, only 196 (92.45%) unique studies out of these 212 studies were covered by the three search strings, culminating in a total literature coverage of merely 74.81% of the library of 262 studies (**Table 8**). Expanding LRN's database integration beyond PubMed could have potentially addressed this shortfall[42].

Increased computational capacity could lead to more efficient processing times, translating into a notable reduction in human labor hours required for SLR completion with LRN (**Table 10**). This efficiency not only streamlines the review process but also has the potential to lower associated costs. A thorough cost-benefit analysis of LRN's implementation is essential to fully understand its impact on productivity and financial savings. Coupled with this cost-benefit analysis would be an assessment of LRN's completeness and interpretability to validate the responses (e.g., complete SLRs) generated by LRN[43]. The substantial decrease in human labor time from 19,920 minutes, or 332 hours over 11 months for the manual SLR to 288.6 minutes, or 4.81 hours over the span of a work week (5 days) for the LRN models exemplified the efficiency and productivity offered by employing XAI for SLRs.

## 5. Conclusions

LRN's performance evinced that direct training by an SME was not imperative for it to achieve deep and accurate text classifications equivalent to an SME in surgical gloving practices. LRN is the first AI platform to offer explainability in tandem with text summarization as a core feature. Moreover, LRN as an XAI platform accurately and reliably demonstrated its ability to conduct SLRs while conforming to PRISMA 2020 guidelines. After 4.81 hours of human labor time, one highly accurate LRN model covered 91.51% of the accessible literature, and 74.81% of the entire corpus, from a SME-curated SLR library.

A limitation in this study was the assumption that the current SME corpus is the ground truth. While this assumption served valuable for the purpose of this study's aims, future validation of the LRN platform will require a prospective study. In a prospective study, human SMEs would need to review the results of LRN *a priori* to conducting a SLR. The UMLS controlled vocabulary system enabled LRN to map different concepts to unique identifiers and allowed cross-walking between different vocabulary systems such as MeSH and ICD-10. The LRN codebase should also be updated to improve automation efficiencies and to handle Russian and Chinese languages.

LRN effectively streamlined SLR methodologies without sacrificing the scientific rigor necessary to achieve the high quality of evidence expected of SLRs. By offering a more accurate, precise, and transparent approach to conducting SLRs in healthcare, LRN's current trajectory suggests significant advancements in automating and optimizing research. Moreover, LRN's versatility allows it to encompass many fields within the life sciences and to generate other research such as meta-analyses, scoping reviews, accelerate method development, and produce evidence summaries. Future efforts will focus on expanding the number of databases LRN can access, paving the way for more comprehensive and up-to-date evidence syntheses in surgical practice and beyond.

## Figures

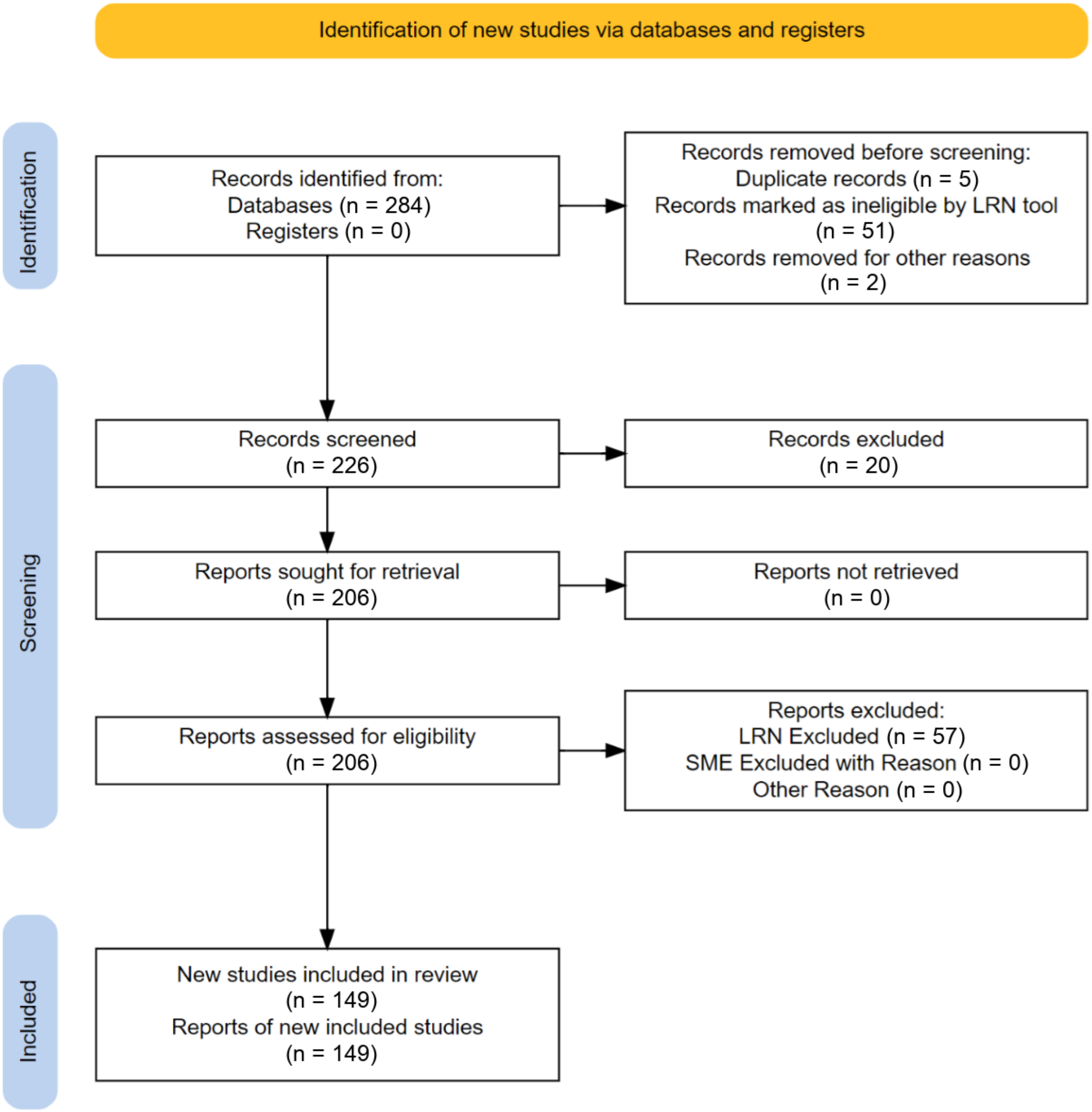

**Figure 1 PRISMA 2020 flow diagram generated by LRN tracking data utilized to train and validate the Highest Performance LRN model**. This flow diagram was automatically produced by LRN upon model finalization for the LRN model with the highest Cohen's kappa. Data source was exclusively PubMed. "Records removed for other reasons" were those records excluded due to language restrictions, and "records deemed ineligible were based on the LRN translated exclusion criteria. "Records excluded" (n = 20) were records excluded during the human user screening phase. Preferred Reporting Items for Systematic Reviews and Meta-Analysis (PRISMA).

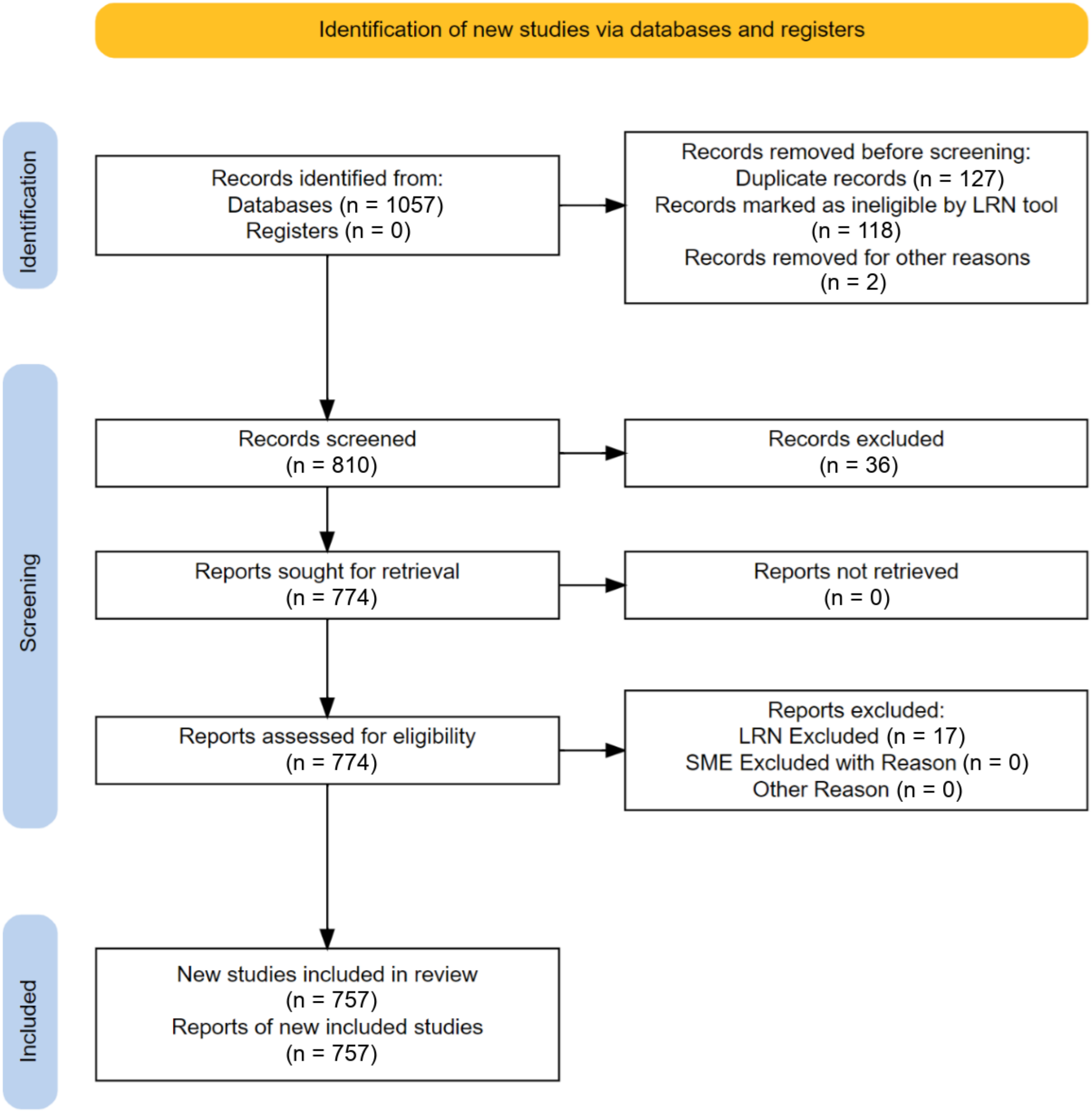

**Figure 2. PRISMA 2020 flow diagram by LRN detailing a systematic literature review for the entire surgical glove corpus with the Optimally Balanced LRN model.** A flow diagram auto-generated post-model finalization, with data sourced from PubMed. The optimally balanced LRN model was derived from search string 1 iteration 2; this deployed model was applied to the entire corpus from the 3 search strings. Exclusions were due to language filtration; ineligibility was based on the combined LRN translated exclusion criteria from 3 search strings. Abstracts with identifiable information (authors, PMIDs, publication date, key words) for the remaining 810 records were exported by LRN to the user upon model finalization. Preferred Reporting Items for Systematic Reviews and Meta-Analysis (PRISMA).

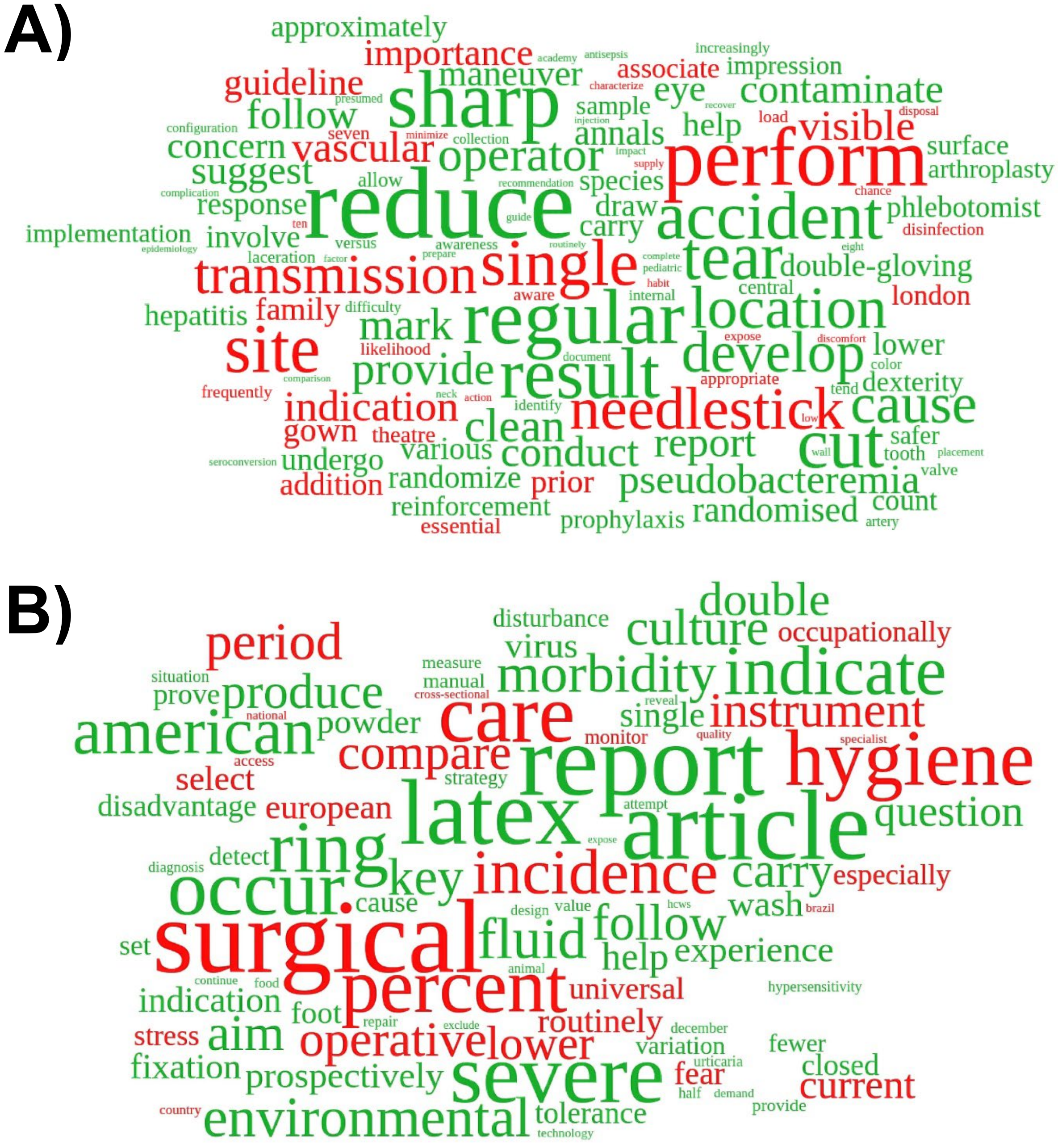

**Figure 3. Tag Clouds from the final iteration of two LRN models reveals key and novel insights into surgical gloving practices.** These visualizations highlight associations identified by LRN within the literature, capturing both expected concepts aligned with SME perspectives and novel insights. These concepts include numerical values and measures, phrases, and acronyms, which were tagged utilizing LRN's word embedding model. Size of the tag relates to the term frequency. Colors indicate relevance to classification: green for INCLUDE and red for EXCLUDE. Figure 3A = highest performance LRN model from search string 3, iteration 3; Figure 3B = optimally balanced LRN model produced by search string 1, iteration 2.

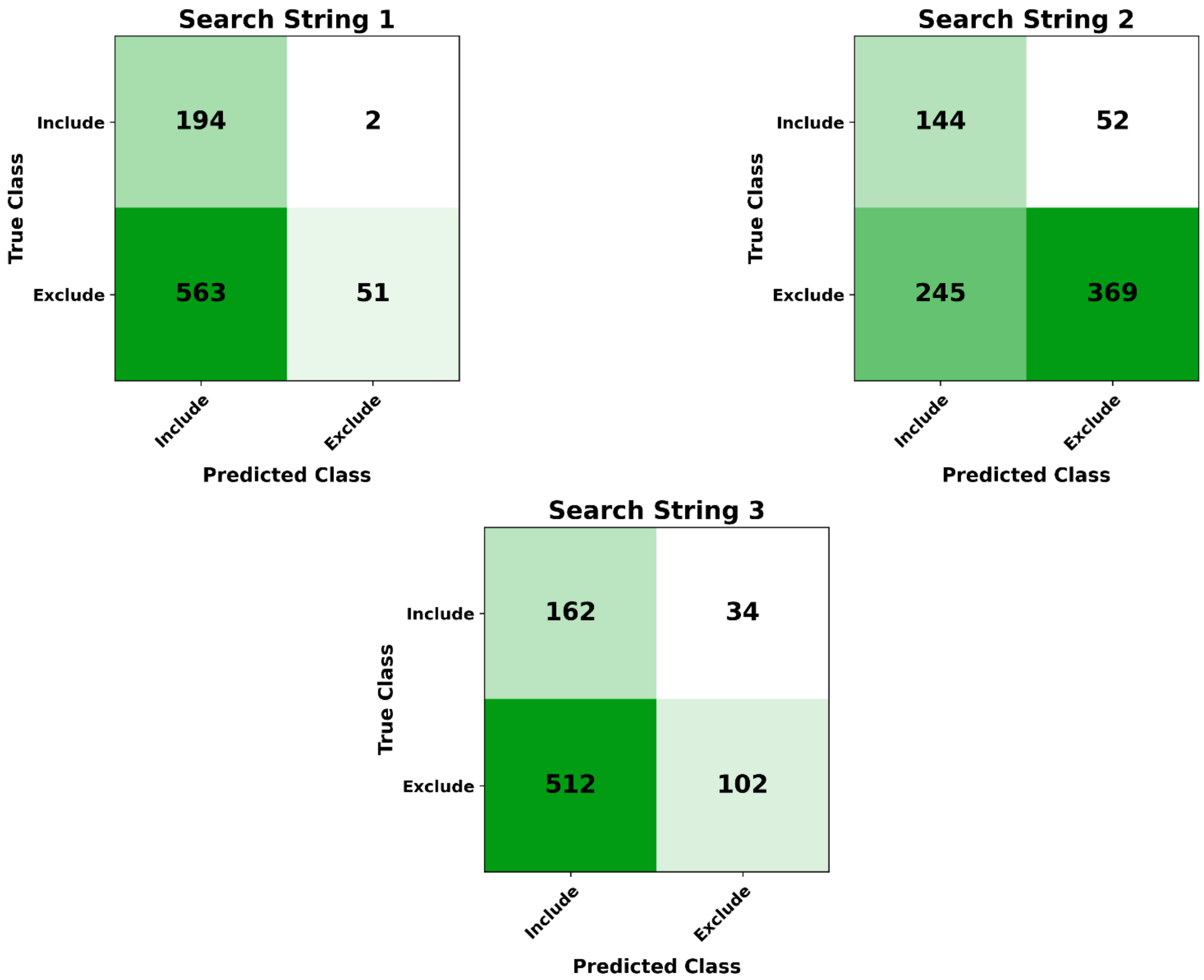

**Figure 4. Confusion matrices comparing separate LRN models' performance in text classification for surgical gloving practices.** Classification outcomes of three LRN models, with optimal models from each search string. Total corpus was 810 full-text reports on surgical gloving practices. Search string 1 (iteration 2) was the optimally balanced LRN model, demonstrating superior ability to accurately classify the highest number of true include articles with the fewest false excludes (negatives). Highest performance LRN model (search string 3 iteration 3) identified more true exclude reports at the expense of misidentifying relatively more include reports. Comparatively, search string 2 (iteration 3) underperformed with the highest number of false excludes.

# Tables

## Table 1. Search strategy configuration for LRN systematic literature review of surgical glove practices.

| Search String Number | Search String | Translated Query | Exclusion Search String | Translated Exclusion Query |
|---|---|---|---|---|
| **1** | ((surgical glove))) AND (((change))) AND (1980/01/01:2023/01/01[dp]) | ("gloves, surgical"[MeSH Terms] OR ("gloves"[All Fields] AND "surgical"[All Fields]) OR "surgical gloves"[All Fields] OR ("surgical"[All Fields] AND "glove"[All Fields]) OR "surgical glove"[All Fields]) AND ("change"[All Fields] OR "changed"[All Fields] OR "changes"[All Fields] OR "changing"[All Fields] OR "changings"[All Fields]) AND 1980/01/01:2023/01/01[Date - Publication] | (((surgical glove))) AND (((change))) AND (1980/01/01:2023/01/01[dp]) AND (((dentistry) OR (orthodontics) OR (veterinary surgery) OR (veterinary))) | ("gloves, surgical"[MeSH Terms] OR ("gloves"[All Fields] AND "surgical"[All Fields]) OR "surgical gloves"[All Fields] OR ("surgical"[All Fields] AND "glove"[All Fields]) OR "surgical glove"[All Fields]) AND ("change"[All Fields] OR "changed"[All Fields] OR "changes"[All Fields] OR "changing"[All Fields] OR "changings"[All Fields]) AND 1980/01/01:2023/01/01[Date - Publication] AND ("dentistry"[MeSH Terms] OR "dentistry"[All Fields] OR "dentistry s"[All Fields] OR ("orthodontal"[All Fields] OR "orthodontic"[All Fields] OR "orthodontical"[All Fields] OR "orthodontically"[All Fields] OR "orthodontics"[MeSH Terms] OR "orthodontics"[All Fields]) OR ("surgery, veterinary"[MeSH Terms] OR ("surgery"[All Fields] AND "veterinary"[All Fields]) OR "veterinary surgery"[All Fields] OR ("veterinary"[All Fields] AND "surgery"[All Fields])) OR ("veterinary"[MeSH Subheading] OR "veterinary"[All Fields])) |
| **2** | (((surgical glove))) AND (((perforation))) AND (1980/01/01:2023/01/01[dp]) | ("gloves, surgical"[MeSH Terms] OR ("gloves"[All Fields] AND "surgical"[All Fields]) OR "surgical gloves"[All Fields] OR ("surgical"[All Fields] AND "glove"[All Fields]) OR "surgical glove"[All Fields]) AND ("perforant"[All Fields] OR "perforants"[All Fields] OR "perforate"[All Fields] OR "perforated"[All Fields] OR "perforates"[All Fields] OR "perforating"[All Fields] OR "perforation"[All Fields] OR "perforations"[All Fields] OR "perforative"[All Fields] OR "perforator"[All Fields] OR "perforator s"[All Fields] OR "perforators"[All Fields]) AND 1980/01/01:2023/01/01[Date - Publication] | (((surgical glove))) AND (((perforation))) AND (1980/01/01:2023/01/01[dp]) AND (((dentistry) OR (orthodontics) OR (veterinary surgery) OR (veterinary))) | ("gloves, surgical"[MeSH Terms] OR ("gloves"[All Fields] AND "surgical"[All Fields]) OR "surgical gloves"[All Fields] OR ("surgical"[All Fields] AND "glove"[All Fields]) OR "surgical glove"[All Fields]) AND ("perforant"[All Fields] OR "perforants"[All Fields] OR "perforate"[All Fields] OR "perforated"[All Fields] OR "perforates"[All Fields] OR "perforating"[All Fields] OR "perforation"[All Fields] OR "perforations"[All Fields] OR "perforative"[All Fields] OR "perforator"[All Fields] OR "perforator s"[All Fields] OR "perforators"[All Fields]) AND 1980/01/01:2023/01/01[Date - Publication] AND ("dentistry"[MeSH Terms] OR "dentistry"[All Fields] OR "dentistry s"[All Fields] OR ("orthodontal"[All Fields] OR "orthodontic"[All Fields] OR "orthodontical"[All Fields] OR "orthodontically"[All Fields] OR "orthodontics"[MeSH Terms] OR "orthodontics"[All Fields]) OR ("surgery, veterinary"[MeSH Terms] OR ("surgery"[All Fields] AND "veterinary"[All Fields]) OR "veterinary surgery"[All Fields] OR ("veterinary"[All Fields] AND "surgery"[All Fields])) OR ("veterinary"[MeSH Subheading] OR "veterinary"[All Fields])) |
| **3** | (((surgery and glove) OR (surgical glove))) AND (((puncture))) AND (1980/01/01:2023/01/01[dp]) | ((("surgery"[MeSH Subheading] OR "surgery"[All Fields] OR "surgical procedures, operative"[MeSH Terms] OR ("surgical"[All Fields] AND "procedures"[All Fields] AND "operative"[All Fields]) OR "operative surgical procedures"[All Fields] OR "general surgery"[MeSH Terms] OR ("general"[All Fields] AND "surgery"[All Fields]) OR "general surgery"[All Fields] OR "surgery s"[All Fields] OR "surgerys"[All Fields] OR "surgeries"[All Fields]) AND ("glove s"[All Fields] OR "gloved"[All Fields] OR "gloves, protective"[MeSH Terms] OR ("gloves"[All Fields] AND "protective"[All Fields]) OR "protective gloves"[All Fields] OR "glove"[All Fields] OR "gloves"[All Fields] OR "gloving"[All Fields])) OR ("gloves, surgical"[MeSH Terms] OR ("gloves"[All Fields] AND "surgical"[All Fields]) OR "surgical gloves"[All Fields] OR ("surgical"[All Fields] AND "glove"[All Fields]) OR "surgical glove"[All Fields])) AND ("punctured"[All Fields] OR "punctures"[MeSH Terms] OR "punctures"[All Fields] OR "puncture"[All Fields] OR "puncturing"[All Fields]) AND 1980/01/01:2023/01/01[Date - Publication] | (((surgery and glove) OR (surgical glove))) AND (((puncture))) AND (1980/01/01:2023/01/01[dp]) AND (((dentistry) OR (orthodontics) OR (veterinary surgery) OR (veterinary))) | ((("surgery"[MeSH Subheading] OR "surgery"[All Fields] OR "surgical procedures, operative"[MeSH Terms] OR ("surgical"[All Fields] AND "procedures"[All Fields] AND "operative"[All Fields]) OR "operative surgical procedures"[All Fields] OR "general surgery"[MeSH Terms] OR ("general"[All Fields] AND "surgery"[All Fields]) OR "general surgery"[All Fields] OR "surgery s"[All Fields] OR "surgerys"[All Fields] OR "surgeries"[All Fields]) AND ("glove s"[All Fields] OR "gloved"[All Fields] OR "gloves, protective"[MeSH Terms] OR ("gloves"[All Fields] AND "protective"[All Fields]) OR "protective gloves"[All Fields] OR "glove"[All Fields] OR "gloves"[All Fields] OR "gloving"[All Fields])) OR ("gloves, surgical"[MeSH Terms] OR ("gloves"[All Fields] AND "surgical"[All Fields]) OR "surgical gloves"[All Fields] OR ("surgical"[All Fields] AND "glove"[All Fields]) OR "surgical glove"[All Fields])) AND ("punctured"[All Fields] OR "punctures"[MeSH Terms] OR "punctures"[All Fields] OR "puncture"[All Fields] OR "puncturing"[All Fields]) AND 1980/01/01:2023/01/01[Date - Publication] AND ("dentistry"[MeSH Terms] OR "dentistry"[All Fields] OR "dentistry s"[All Fields] OR ("orthodontal"[All Fields] OR "orthodontic"[All Fields] OR "orthodontical"[All Fields] OR "orthodontically"[All Fields] OR "orthodontics"[MeSH Terms] OR "orthodontics"[All Fields]) OR ("surgery, veterinary"[MeSH Terms] OR ("surgery"[All Fields] AND "veterinary"[All Fields]) OR "veterinary surgery"[All Fields] OR ("veterinary"[All Fields] AND "surgery"[All Fields])) OR ("veterinary"[MeSH Subheading] OR "veterinary"[All Fields])) |

**Legend:** Three separate search strategies employed through three separate LRN models. Adapted from PICOT used in manual literature reviews. Search strings were the original search strings, while exclusion search strings were the original search strings AND exclusion criteria. Translated queries were those inputted in LRN that were automatically converted into MeSH terms for PubMed REST API calls.

**Table 2. Transparent, iteration-wise evolution of user-defined concept rules across search strings.**

| Rule Number | String 1 Rule | String 1 Label | Iteration Modified | String 2 Rule | String 2 Label | Iteration Modified | String 3 Rule | String 3 Label | Iteration Modified |
|---|---|---|---|---|---|---|---|---|---|
| 1 | contamination | INCLUDE | 1 | contamination | INCLUDE | 1 | contamination | INCLUDE | 1 |
| 2 | latex | INCLUDE | 1 | latex | INCLUDE | 1 | latex | INCLUDE | 1 |
| 3 | polyisoprene | INCLUDE | 1 | polyisoprene | INCLUDE | 1 | polyisoprene | INCLUDE | 1 |
| 4 | polychloroprene | INCLUDE | 1 | polychloroprene | INCLUDE | 1 | polychloroprene | INCLUDE | 1 |
| 5 | procedural | INCLUDE | 1 | procedural | INCLUDE | 1 / 4 | procedural | INCLUDE | 1 |
| 6 | glove | INCLUDE | 1 | glove | INCLUDE | 1 | glove | INCLUDE | 1 |
| 7 | operation | INCLUDE | 1 | operation | INCLUDE | 1 | operation | INCLUDE | 1 |
| 8 | puncture | INCLUDE | 1 | puncture | INCLUDE | 1 | puncture | INCLUDE | 1 |
| 9 | perioperative | INCLUDE | 1 | perioperative | INCLUDE | 1 | perioperative | INCLUDE | 1 |
| 10 | perforation | INCLUDE | 1 | perforation | INCLUDE | 1 | perforation | INCLUDE | 1 |
| 11 | experiment | EXCLUDE | 1 | experiment | EXCLUDE | 1 | experiment | EXCLUDE | 1 |
| 12 | clean glove | EXCLUDE | 1 | clean glove | EXCLUDE | 1 | clean glove | EXCLUDE | 1 |
| 13 | exam glove | EXCLUDE | 1 | exam glove | EXCLUDE | 1 | exam glove | EXCLUDE | 1 |
| 14 | vinyl | EXCLUDE | 1 | vinyl | EXCLUDE | 1 | vinyl | EXCLUDE | 1 |
| 15 | nitrile | EXCLUDE | 1 | nitrile | EXCLUDE | 1 | nitrile | EXCLUDE | 1 |
| 16 | condom | EXCLUDE | 2 | condom | EXCLUDE | 2 | condom | EXCLUDE | 2 |
| 17 | animal | EXCLUDE | 2 | maxillofacial | EXCLUDE | 2 | penetration | INCLUDE | 2 |
| 18 | wash | EXCLUDE | 2 / 3 | double gloving | INCLUDE | 2 | needle penetration | INCLUDE | 2 |
| 19 | surgical gloves | INCLUDE | 2 | perforations | INCLUDE | 2 | breach | INCLUDE | 2 |
| 20 | talc | EXCLUDE | 2 | indication system | INCLUDE | 2 | hole | INCLUDE | 2 |
| 21 | laboratory | EXCLUDE | 2 | intermaxillary | EXCLUDE | 2 | surgical gloves | INCLUDE | 2 |
| 22 | dentist | EXCLUDE | 2, 4 / 3 | arthroplasty | INCLUDE | 2 | allergic | INCLUDE | 2 |
| 23 | animals | EXCLUDE | 2 | surgical gloves | INCLUDE | 2 | barrier | INCLUDE | 2 |
| 24 | antibiotic prophylaxis | EXCLUDE | 2 | examine | EXCLUDE | 2 | intra-operative | INCLUDE | 2 |
| 25 | latex glove | INCLUDE | 2 | examination | EXCLUDE | 2 | surgical | INCLUDE | 2 |
| 26 | double gloving | INCLUDE | 2 | examination glove | EXCLUDE | 2 | surgical glove | INCLUDE | 2 |
| 27 | gloving method | INCLUDE | 2 | vinyl glove | EXCLUDE | 2 | examination glove | EXCLUDE | 2 |
| 28 | allergy | EXCLUDE | 2 / 3 | nitrile glove | EXCLUDE | 2 | examination gloves | EXCLUDE | 2 |
| 29 | vinyl glove | EXCLUDE | 2 | barrier | INCLUDE | 2 | double glove | INCLUDE | 2 |
| 30 | dental | EXCLUDE | 2, 4 / 3 | double-gloving | INCLUDE | 2 | single glove | INCLUDE | 2 |
| 31 | silicone | EXCLUDE | 2 | animal | EXCLUDE | 2 | punctures | INCLUDE | 2 |
| 32 | solution | EXCLUDE | 2 | dentistry | EXCLUDE | 2 | latex gloves | INCLUDE | 2 |
| 33 | allergenic | EXCLUDE | 2 | veterinary | EXCLUDE | 2 | double-gloving | INCLUDE | 2 |
| 34 | culture | EXCLUDE | 2 | orthodontic | EXCLUDE | 2 / 4 | latex glove | INCLUDE | 2 |
| 35 | antiseptic | EXCLUDE | 2 | mandibular | EXCLUDE | 2 | hand washing | EXCLUDE | 2 |
| 36 | disinfectant | EXCLUDE | 2 | rat | EXCLUDE | 2 / 4 | washing | EXCLUDE | 2 |
| 37 | chlorhexidine | EXCLUDE | 2 | mouse | EXCLUDE | 2 / 4 | penetrations | INCLUDE | 2 |
| 38 | changing gloves | INCLUDE | 2 | in vitro | EXCLUDE | 3 | cultures | EXCLUDE | 2 |
| 39 | dentists | EXCLUDE | 2 | soap | EXCLUDE | 3 | blood cultures | EXCLUDE | 2 |
| 40 | glove perforations | INCLUDE | 2 | dental | EXCLUDE | 3 / 4 | culture | EXCLUDE | 2 |
| 41 | immunization | EXCLUDE | 2 | dentists | EXCLUDE | 3 | washed | EXCLUDE | 2 |
| 42 | examination gloves | EXCLUDE | 2 | powder | EXCLUDE | 3 | gloves | INCLUDE | 2 |
| 43 | examination glove | EXCLUDE | 2 | non-latex | EXCLUDE | 3 | hospital | INCLUDE | 2 |
| 44 | latex examination | EXCLUDE | 2 | scrub nurse | INCLUDE | 3 | hand wash | EXCLUDE | 2 |
| 45 | glove change | INCLUDE | 2 | scrub | INCLUDE | 3 / 4 | puncture-resistant | INCLUDE | 3 |
| 46 | nitrile examination | EXCLUDE | 2 / 3 | glove powder | EXCLUDE | 3 | laboratory | EXCLUDE | 3 |
| 47 | nitrile glove | EXCLUDE | 2 | detergent | EXCLUDE | 3 | rat | EXCLUDE | 3 |
| 48 | hand wash | EXCLUDE | 2 | wash | EXCLUDE | 4 | bacterial | EXCLUDE | 3 |
| 49 | hand washing | EXCLUDE | 2 | clinical | INCLUDE | 4 | - | - | - |
| 50 | gloved | INCLUDE | 2 / 3 | gloved | INCLUDE | 4 | - | - | - |
| 51 | double glove | INCLUDE | 2 | study | INCLUDE | 4 | - | - | - |
| 52 | double-gloving | INCLUDE | 2 / 3 | double gloves | INCLUDE | 4 | - | - | - |
| 53 | single glove | INCLUDE | 2 | gloving | INCLUDE | 4 | - | - | - |
| 54 | hand rub | EXCLUDE | 3 | change | INCLUDE | 4 | - | - | - |
| 55 | washing | EXCLUDE | 3 | gloves | INCLUDE | 4 | - | - | - |
| 56 | hospital | INCLUDE | 3 | postoperative | INCLUDE | 4 | - | - | - |
| 57 | damage | INCLUDE | 3 | glove reinforcement | INCLUDE | 4 | - | - | - |
| 58 | asepsis | INCLUDE | 3 | double | INCLUDE | 4 | - | - | - |
| 59 | surgery | INCLUDE | 3 | post mortem | EXCLUDE | 4 | - | - | - |
| 60 | single-gloving | INCLUDE | 4 | hand washed | EXCLUDE | 4 | - | - | - |
| 61 | change gloves | INCLUDE | 4 | procedures | INCLUDE | 4 | - | - | - |
| 62 | alcohol-based | EXCLUDE | 4 | surgical procedure | INCLUDE | 4 | - | - | - |
| 63 | protective | INCLUDE | 4 | perforated | INCLUDE | 4 | - | - | - |
| 64 | barrier | INCLUDE | 4 | penetration | INCLUDE | 4 | - | - | - |
| 65 | operating room | INCLUDE | 4 | - | - | - | - | - | - |
| 66 | operating theater | INCLUDE | 4 | - | - | - | - | - | - |
| 67 | vinyl gloves | EXCLUDE | 4 | - | - | - | - | - | - |
| 68 | breach | INCLUDE | 4 | - | - | - | - | - | - |

**Legend:** LRN tracks decisions made by the user during screening by tracking each rule within the search strings, which are designated by user-assigned labels (INCLUDE or EXCLUDE). Each entry shows the iteration in which a rule was added, and, if applicable, subsequent iterations where it was removed and/or reinstated. For instance, for the rule "dentist" under search string 1, the '2,4 / 3' indicates the rule was initially added in iteration 2, removed in iteration 3, and reinstated in iteration 4. Rules can take the form of numerical values, acronyms, terms, or phrases.

**Table 3. User submitted prompts to LRN for systematic literature review generation.**

| Section | Prompt | Generation |
|---|---|---|
| **Introduction** | "What are the gaps in understanding the relationship between glove damage and the frequency of glove changes during various surgical procedures?" | 1 |
| **Results** | "What is the relationship between glove damage and the frequency of glove changes recommended during various types of surgical procedures? Are there established guidelines or specialty-specific recommendations for changing gloves during surgery to minimize the risk of glove damage and maintain sterility? How does the incidence of glove damage correlate with the duration of surgical procedures, and what are the current best practices for glove change frequency to ensure patient and healthcare worker safety?" | 1 |
| **Discussion** | "How do the results of these studies contribute to the understanding of the optimal strategies for glove change frequency to minimize glove damage, and based on these studies what gaps or challenges remain in the field?" | 1 |

**Legend:** LRN requires the user to submit at minimum 3 questions per generation of a systematic literature review (SLR), or other data product (e.g., a meta-analysis). For an SLR, prompts are divided into three sections: an introduction, results, and discussion. Prompts may contain single or multiple questions, direct instructions for desired contexts, and do not require proper grammar. A generation represents one version of the SLR produced by a set of questions submitted to LRN.

**Table 4. Iterative performance metrics of LRN models across search strings.**

| Search String | Cohen's Kappa | Accuracy | Average Potential |
|---|---|---|---|
| **Iteration 1** | | | |
| String 1 | 0.1348 | 86.90% | 33.62% |
| String 2 | 0.0000 | 89.66% | 86.39% |
| String 3 | 0.0000 | 88.89% | 85.41% |
| **Iteration 2** | | | |
| ***String 1†*** | ***0.2174†*** | ***85.71%†*** | ***43.99%†*** |
| String 2 | 0.0000 | 87.95% | 83.39% |
| String 3 | 0.0000 | 84.78% | 85.44% |
| **Iteration 3** | | | |
| String 1 | 0.0000 | 82.14% | 85.06% |
| String 2 | 0.0183 | 58.62% | 53.59% |
| ***String 3**** | ***0.4953**** | ***84.78%**** | ***49.28%**** |

**Legend:** LRN models are configured with the goal to maximize the performance metrics—Cohen's kappa, overall accuracy, and average potential—for each search string across four iterations. The optimally balanced model (†) was search string 1 iteration 2, and the highest performance model, search string 3 iteration 3, is denoted by an asterisk (). Metrics reflect the integration of data up to the third iteration for each model, showcasing the cumulative improvement in performance. The highest performance model (*) was distinguished by its superior Cohen's kappa and high overall accuracy. Per iteration, the titles and abstracts of 20 records were presented to the user for feedback.

**Table 5. Individual class metrics for highest performance LRN model.**

| Label | Recall | Precision | F-score |
|---|---|---|---|
| **Iteration 1** | | | |
| INCLUDE | 88.89% | 100.00% | 94.12% |
| EXCLUDE | 0.00% | 0.00% | 0.00% |
| **Iteration 2** | | | |
| INCLUDE | 84.78% | 88.89% | 88.89% |
| EXCLUDE | 0.00% | 0.00% | 0.00% |
| **Iteration 3** | | | |
| INCLUDE | 91.89% | 89.47% | 90.67% |
| EXCLUDE | 55.56% | 62.50% | 58.82% |

**Legend:** LRN model with the highest performance, identified in iteration 3 of search string 3. Improvements in recall and precision for the EXCLUDE label at the expense of initially reduced INCLUDE precision.

**Table 6. Individual class metrics for optimally balanced LRN model.**

| Label | Recall | Precision | F-score |
|---|---|---|---|
| **Iteration 1** | | | |
| INCLUDE | 100.00% | 86.75% | 92.90% |
| EXCLUDE | 8.33% | 100.00% | 15.38% |
| **Iteration 2** | | | |
| INCLUDE | 100.00% | 85.37% | 92.11% |
| EXCLUDE | 14.29% | 100.00% | 25.00% |
| **Iteration 3** | | | |
| INCLUDE | 100.00% | 82.14% | 90.20% |
| EXCLUDE | 0.00% | 0.00% | 0.00% |

**Legend:** LRN model with the highest number of true positives, denoted the optimally balanced LRN model, identified in iteration 2 of search string 1. Minor improvements in EXCLUDE recall with marginal reduction of INCLUDE precision after one RLHF iteration.

**Table 7. Significant concept rules guiding decision-making by the highest performance LRN model.**

| Rule 1 | Rule 2 | Rule 1 Class | Rule 2 Class | Correlation value | P-value (raw) | P-value (FDR-adjusted) | Rule 1 Report Coverage | Rule 2 Report Coverage |
|---|---|---|---|---|---|---|---|---|
| nitrile | examination glove | EXCLUDE | EXCLUDE | 0.6009 | 1.170E-14 | 5.302E-13 | 9 / 226 | 10 / 226 |
| nitrile | examination gloves | EXCLUDE | EXCLUDE | 0.6009 | 1.170E-14 | 5.302E-13 | 9 / 226 | 10 / 226 |
| condom | hand washing | EXCLUDE | EXCLUDE | 0.5052 | 8.631E-11 | 3.716E-09 | 4 / 226 | 6 / 226 |
| polychloroprene | nitrile | INCLUDE | EXCLUDE | 0.4954 | 1.971E-10 | 8.080E-09 | 1 / 226 | 9 / 226 |
| operation | surgical gloves | INCLUDE | INCLUDE | 0.4901 | 3.054E-10 | 1.052E-08 | 136 / 226 | 69 / 226 |
| operation | surgical glove | INCLUDE | INCLUDE | 0.4901 | 3.054E-10 | 1.052E-08 | 136 / 226 | 69 / 226 |
| surgical gloves | surgical | INCLUDE | INCLUDE | 0.4901 | 3.054E-10 | 1.052E-08 | 69 / 226 | 136 / 226 |
| surgical | surgical glove | INCLUDE | INCLUDE | 0.4901 | 3.054E-10 | 1.052E-08 | 136 / 226 | 69 / 226 |
| contamination | cultures | INCLUDE | EXCLUDE | 0.4300 | 3.327E-08 | 1.061E-06 | 38 / 226 | 19 / 226 |
| contamination | culture | INCLUDE | EXCLUDE | 0.4300 | 3.327E-08 | 1.061E-06 | 38 / 226 | 19 / 226 |
| polychloroprene | exam glove | INCLUDE | EXCLUDE | 0.4020 | 2.424E-07 | 6.957E-06 | 1 / 226 | 6 / 226 |
| polychloroprene | examination glove | INCLUDE | EXCLUDE | 0.4020 | 2.424E-07 | 6.957E-06 | 1 / 226 | 10 / 226 |
| polychloroprene | examination gloves | INCLUDE | EXCLUDE | 0.4020 | 2.424E-07 | 6.957E-06 | 1 / 226 | 10 / 226 |
| contamination | blood cultures | INCLUDE | EXCLUDE | 0.3946 | 4.011E-07 | 1.114E-05 | 38 / 226 | 10 / 226 |
| exam glove | nitrile | EXCLUDE | EXCLUDE | 0.3904 | 5.311E-07 | 1.429E-05 | 6 / 226 | 9 / 226 |
| nitrile | latex glove | EXCLUDE | INCLUDE | 0.3644 | 2.849E-06 | 7.432E-05 | 9 / 226 | 34 / 226 |
| latex | double glove | INCLUDE | INCLUDE | 0.3611 | 3.520E-06 | 8.660E-05 | 49 / 226 | 40 / 226 |
| double glove | latex gloves | INCLUDE | INCLUDE | 0.3611 | 3.520E-06 | 8.660E-05 | 40 / 226 | 49 / 226 |
| double glove | single glove | INCLUDE | INCLUDE | 0.3573 | 4.451E-06 | 0.0001 | 40 / 226 | 17 / 226 |
| surgical gloves | double glove | INCLUDE | INCLUDE | 0.3521 | 6.086E-06 | 0.0001 | 69 / 226 | 40 / 226 |
| surgical glove | double glove | INCLUDE | INCLUDE | 0.3521 | 6.086E-06 | 0.0001 | 69 / 226 | 40 / 226 |
| puncture | blood cultures | INCLUDE | EXCLUDE | 0.3490 | 7.339E-06 | 0.0002 | 167 / 226 | 10 / 226 |
| punctures | blood cultures | INCLUDE | EXCLUDE | 0.3490 | 7.339E-06 | 0.0002 | 167 / 226 | 10 / 226 |
| vinyl | nitrile | EXCLUDE | EXCLUDE | 0.3426 | 1.076E-05 | 0.0002 | 4 / 226 | 9 / 226 |
| operation | puncture | INCLUDE | INCLUDE | 0.3356 | 1.622E-05 | 0.0003 | 136 / 226 | 167 / 226 |
| operation | punctures | INCLUDE | INCLUDE | 0.3356 | 1.622E-05 | 0.0003 | 136 / 226 | 167 / 226 |
| puncture | surgical | INCLUDE | INCLUDE | 0.3356 | 1.622E-05 | 0.0003 | 167 / 226 | 136 / 226 |
| surgical | punctures | INCLUDE | INCLUDE | 0.3356 | 1.622E-05 | 0.0003 | 136 / 226 | 167 / 226 |
| perforation | surgical gloves | INCLUDE | INCLUDE | 0.3279 | 2.526E-05 | 0.0005 | 58 / 226 | 69 / 226 |
| perforation | surgical glove | INCLUDE | INCLUDE | 0.3279 | 2.526E-05 | 0.0005 | 58 / 226 | 69 / 226 |
| contamination | bacterial | INCLUDE | EXCLUDE | 0.3160 | 4.922E-05 | 0.0009 | 38 / 226 | 18 / 226 |
| penetration | needle penetration | INCLUDE | INCLUDE | 0.3158 | 4.978E-05 | 0.0009 | 14 / 226 | 58 / 226 |
| needle penetration | penetrations | INCLUDE | INCLUDE | 0.3158 | 4.978E-05 | 0.0009 | 58 / 226 | 14 / 226 |
| hole | washing | INCLUDE | EXCLUDE | 0.3122 | 6.080E-05 | 0.0010 | 22 / 226 | 9 / 226 |

**Legend:** Significantly correlated concepts were those with strong evidence (FDR-adjusted P-value < 0.001). FDR = false discovery rate (Benjamini-Hochberg method). Training and validation set consisted of 226 records screened by LRN for search string 3 iteration 3. Normalized chi-square values with Cramer's V constrained values into a range of [0,1].

**Table 8. Significant concept rules influencing text classification by the optimally balanced LRN model.**

| Rule 1 | Rule 2 | Rule 1 Class | Rule 2 Class | Correlation value | P-value (raw) | P-value (FDR-adjusted) | Rule 1 Report Coverage | Rule 2 Report Coverage |
|---|---|---|---|---|---|---|---|---|
| talc | animals | EXCLUDE | EXCLUDE | 0.4041 | 6.120E-14 | 2.424E-12 | 3 / 417 | 6 / 417 |
| animal | talc | EXCLUDE | EXCLUDE | 0.4041 | 6.120E-14 | 2.424E-12 | 6 / 417 | 3 / 417 |
| condom | antibiotic prophylaxis | EXCLUDE | EXCLUDE | 0.3969 | 1.687E-13 | 6.424E-12 | 14 / 417 | 42 / 417 |
| exam glove | examination glove | EXCLUDE | EXCLUDE | 0.3685 | 7.665E-12 | 2.710E-10 | 4 / 417 | 14 / 417 |
| exam glove | examination gloves | EXCLUDE | EXCLUDE | 0.3685 | 7.665E-12 | 2.710E-10 | 4 / 417 | 14 / 417 |
| operation | surgical gloves | INCLUDE | INCLUDE | 0.3672 | 9.047E-12 | 3.088E-10 | 283 / 417 | 116 / 417 |
| latex glove | allergenic | INCLUDE | EXCLUDE | 0.3636 | 1.434E-11 | 4.734E-10 | 48 / 417 | 13 / 417 |
| examination gloves | latex examination | EXCLUDE | EXCLUDE | 0.3496 | 8.430E-11 | 2.608E-09 | 14 / 417 | 60 / 417 |
| examination glove | latex examination | EXCLUDE | EXCLUDE | 0.3496 | 8.430E-11 | 2.608E-09 | 14 / 417 | 60 / 417 |
| contamination | culture | INCLUDE | EXCLUDE | 0.3460 | 1.303E-10 | 3.909E-09 | 62 / 417 | 29 / 417 |
| latex | allergenic | INCLUDE | EXCLUDE | 0.3312 | 7.681E-10 | 2.236E-08 | 83 / 417 | 13 / 417 |
| nitrile | examination glove | EXCLUDE | EXCLUDE | 0.3266 | 1.309E-09 | 3.241E-08 | 7 / 417 | 14 / 417 |
| nitrile | examination gloves | EXCLUDE | EXCLUDE | 0.3266 | 1.309E-09 | 3.241E-08 | 7 / 417 | 14 / 417 |
| examination gloves | nitrile examination | EXCLUDE | EXCLUDE | 0.3266 | 1.309E-09 | 3.241E-08 | 14 / 417 | 7 / 417 |
| examination glove | nitrile examination | EXCLUDE | EXCLUDE | 0.3266 | 1.309E-09 | 3.241E-08 | 14 / 417 | 7 / 417 |
| examination gloves | nitrile glove | EXCLUDE | EXCLUDE | 0.3266 | 1.309E-09 | 3.241E-08 | 14 / 417 | 7 / 417 |
| examination glove | nitrile glove | EXCLUDE | EXCLUDE | 0.3266 | 1.309E-09 | 3.241E-08 | 14 / 417 | 7 / 417 |
| antiseptic | chlorhexidine | EXCLUDE | EXCLUDE | 0.3219 | 2.249E-09 | 5.430E-08 | 21 / 417 | 20 / 417 |
| talc | silicone | EXCLUDE | EXCLUDE | 0.3102 | 8.316E-09 | 1.960E-07 | 3 / 417 | 7 / 417 |
| vinyl | hand washing | EXCLUDE | EXCLUDE | 0.3009 | 2.285E-08 | 5.260E-07 | 13 / 417 | 11 / 417 |
| perforation | surgical gloves | INCLUDE | INCLUDE | 0.2927 | 5.407E-08 | 1.216E-06 | 54 / 417 | 116 / 417 |
| perioperative | condom | INCLUDE | EXCLUDE | 0.2825 | 1.546E-07 | 3.400E-06 | 26 / 417 | 14 / 417 |
| latex | vinyl | INCLUDE | EXCLUDE | 0.2678 | 6.542E-07 | 1.408E-05 | 83 / 417 | 13 / 417 |
| perforation | double glove | INCLUDE | INCLUDE | 0.2671 | 7.028E-07 | 1.450E-05 | 54 / 417 | 51 / 417 |
| perforation | double gloving | INCLUDE | INCLUDE | 0.2671 | 7.028E-07 | 1.450E-05 | 54 / 417 | 51 / 417 |
| exam glove | latex examination | EXCLUDE | EXCLUDE | 0.2631 | 1.027E-06 | 2.075E-05 | 4 / 417 | 60 / 417 |
| perioperative | antibiotic prophylaxis | INCLUDE | EXCLUDE | 0.2588 | 1.535E-06 | 3.039E-05 | 26 / 417 | 42 / 417 |
| double gloving | single glove | INCLUDE | INCLUDE | 0.2569 | 1.828E-06 | 3.480E-05 | 51 / 417 | 14 / 417 |
| double glove | single glove | INCLUDE | INCLUDE | 0.2569 | 1.828E-06 | 3.480E-05 | 51 / 417 | 14 / 417 |
| antibiotic prophylaxis | chlorhexidine | EXCLUDE | EXCLUDE | 0.2505 | 3.266E-06 | 6.100E-05 | 42 / 417 | 20 / 417 |
| animal | silicone | EXCLUDE | EXCLUDE | 0.2499 | 3.465E-06 | 6.237E-05 | 6 / 417 | 7 / 417 |
| animals | silicone | EXCLUDE | EXCLUDE | 0.2499 | 3.465E-06 | 6.237E-05 | 6 / 417 | 7 / 417 |
| solution | chlorhexidine | EXCLUDE | EXCLUDE | 0.2207 | 4.150E-05 | 0.0007 | 25 / 417 | 20 / 417 |
| antibiotic prophylaxis | antiseptic | EXCLUDE | EXCLUDE | 0.2202 | 4.306E-05 | 0.0007 | 42 / 417 | 21 / 417 |

**Legend:** Significantly correlated concepts were those with strong evidence (FDR-adjusted P-value $< 0.001$). FDR = false discovery rate (Benjamini-Hochberg method). Training and validation set consisted of 417 records screened by LRN for search string 1 iteration 2. Normalized chi-square values with Cramer's V constrained values into a range of [0,1].

**Table 9. Jaccard Index analysis for LRN search strings and subject matter expert-curated literature review library.**

| Search String | Comparison Set | Jaccard Index | P-value (raw) | P-value (FDR-adjusted) |
|---|---|---|---|---|
| String 1 | SME Library | 0.2503 | 3.478E-03 | 3.538E-03 |
| String 2 | SME Library | 0.3151 | 2.000E-05 | 3.000E-05 |
| String 3 | SME Library | 0.2238 | 3.538E-03 | 3.538E-03 |
| String 1 | String 2 | 0.5059 | <1.000E-20 | <1.000E-20 |
| String 1 | String 3 | 0.8609 | <1.000E-20 | <1.000E-20 |
| String 2 | String 3 | 0.4682 | <1.000E-20 | <1.000E-20 |

**Legend:** Values indicate similarity levels, with a higher Jaccard index reflecting greater overlap. Significance is determined by p-values adjusted by the false discovery rate (FDR) using the Benjamini-Hochberg method. The highest observed overlap between search strings 1 and 3 suggests substantial consistency in literature coverage. Subject matter expert (SME).

**Table 10. Productivity metrics for human labor versus computational time for LRN literature review processes.**

| String | (Date Start : End) | Human Labor Time (min) | Runtime Start (hr:min:sec) | Runtime End (hr:min:sec) | Total Runtime (hr:min:sec) |
|---|---|---|---|---|---|
| **Iteration 1** | | | | | |
| String 1 | 2023 / 12 / 24 : 2023 / 12 / 24 | 6.57 | 18:48:12 | 22:43:24 | 03:55:12 |
| String 2 | 2023 / 12 / 23 : 2023 / 12 / 23 | 34.03 | 21:50:36 | 23:26:22 | 01:35:46 |
| String 3 | 2023 / 12 / 25 : 2023 / 12 / 25 | 17.42 | 05:27:36 | 06:55:02 | 01:27:26 |
| **Iteration 2** | | | | | |
| String 1 | 2023 / 12 / 25 : 2023 / 12 / 25 | 24.22 | 00:42:37 | 05:37:32 | 04:54:55 |
| String 2 | 2023 / 12 / 24 : 2023 / 12 / 24 | 32.93 | 19:14:34 | 21:06:51 | 01:52:17 |
| String 3 | 2023 / 12 / 26 : 2023 / 12 / 26 | 35.38 | 05:39:34 | 07:25:54 | 01:46:20 |
| **Iteration 3** | | | | | |
| String 1 | 2023 / 12 / 26 : 2023 / 12 / 26 | 14.73 | 17:52:29 | 22:58:06 | 05:05:37 |
| String 2 | 2023 / 12 / 25 : 2023 / 12 / 25 | 41.92 | 00:56:48 | 02:57:08 | 02:00:20 |
| String 3 | 2023 / 12 / 28 : 2023 / 12 / 28 | 30.07 | 00:20:31 | 02:02:57 | 01:42:26 |
| **Iteration 4** | | | | | |
| String 1 | 2023 / 12 / 27 : 2023 / 12 / 28 | 20.97 | 22:17:46 | 01:13:53 | 02:56:07 |
| String 2 | 2023 / 12 / 28 : 2023 / 12 / 28 | 15.90 | 17:42:50 | 19:34:19 | 01:51:29 |
| String 3 | 2023 / 12 / 28 : 2023 / 12 / 28 | 14.47 | 18:26:01 | 19:28:46 | 01:02:45 |
| | **Total Human Labor Time (mins):** | **288.6** | | **Total Computation Time (mins):** | **1810.7** |

**Legend:** Human labor time and computational run times for completing the systematic literature review on surgical glove practices using LRN. Start and end dates, as well as runtime start and runtime end (computation time), are reported in coordinated universal time (UTC) within 24-hour periods. Total human labor time involved configuring the LRN model for iteration 1; subsequent iterations (2+) involved incorporation of user feedback through screening 20 record's abstracts and titles, and through modification of that string's natural language concept ruleset.